\documentclass[12pt,a4paper]{article}
\usepackage{graphicx}
\usepackage{cite}

\makeatletter
\renewcommand{\@biblabel}[1]{}
\makeatother

\begin{document}

\newcommand{\EQ}{Eq.~}
\newcommand{\EQS}{Eqs.~}
\newcommand{\FIG}{Fig.~}
\newcommand{\FIGS}{Figs.~}
\newcommand{\SEC}{Sec.~}
\newcommand{\SECS}{Secs.~}

\setlength{\baselineskip}{0.77cm}

\title{Oscillatory dynamics in evolutionary games are suppressed by
heterogeneous adaptation rates of players}
\bigskip

\author{Naoki Masuda${}^*$\\
\  \\
\ \\
${}^{1}$
Graduate School of Information Science and Technology,\\
The University of Tokyo,\\
7-3-1 Hongo, Bunkyo, Tokyo 113-8656, Japan
\ \\
$^*$ Author for correspondence (masuda@mist.i.u-tokyo.ac.jp)}
\date{}
\maketitle

\newpage

\begin{abstract}
\setlength{\baselineskip}{0.77cm} Game dynamics in which three or more
strategies are cyclically competitive, as represented by the
rock-scissors-paper game, have attracted practical and theoretical
interests. In evolutionary dynamics, cyclic competition results in
oscillatory dynamics of densities of individual strategists. In
finite-size populations, it is known that oscillations blow up until
all but one strategies are eradicated if without mutation.  In the
present paper, we formalize replicator dynamics with players that have
different adaptation rates.  We show analytically and numerically that
the heterogeneous adaptation rate suppresses the oscillation
amplitude.  In social dilemma games with cyclically competing
strategies and homogeneous adaptation rates, altruistic strategies are
often relatively weak and cannot survive in finite-size
populations. In such situations, heterogeneous adaptation rates save
coexistence of different strategies and hence promote altruism. When
one strategy dominates the others without cyclic competition, fast
adaptors earn more than slow adaptors. When not,
mixture of fast and slow adaptors stabilizes population dynamics, and
slow adaptation does not imply inefficiency for a player.
\end{abstract}


\newpage

\section{Introduction}\label{sec:introduction}

Cyclic competition among different phenotypes is often found in
nature.  A minimal system of cyclic competition consists of
individuals of three phenotypes. In so-called rock-scissors-paper
(RSP) dynamics, R beats S, S beats P, and P beats R, hence no 
phenotype is entirely dominant. Examples of natural
RSP dynamics include tropical marine ecosystems \cite{Buss80},
color polymorphism of natural lizards
\cite{Sinervo96}, vertebrate ecosystems in high-arctic areas
\cite{Gilg03}, real microbial communities of {\it Escherichia coli}
\cite{Kerr02}, and various disease dynamics represented by the
susceptible-infected-recovered-susceptible model or its variants
\cite{Andersonbook}.

Also in evolutionary game theory, cyclic competition dynamics realized
by three or more strategies often appear. 
In social dilemma games such
as the Prisoner's Dilemma,
cyclic competition often underlies the survival of altruistic
strategies subject to invasion by 
nonaltruistic strategies.  For example, if the players can take
either unconditional cooperation (ALLC) or unconditional defection
(ALLD) in the Prisoner's Dilemma, ALLD eventually prevails.
However, if discriminators that use reputation of other
players regarding the cooperation tendency are allowed,
ALLC, ALLD, and discriminators form an RSP cycle; ALLD is stronger
than ALLC, ALLC and discriminators are neutral, and discriminators are
stronger than ALLD under certain conditions.  Accordingly, these strategies
alternately prosper in the population, which
implies survival of altruistic players, namely, ALLC and
discriminators \cite{Nowak98NAT,Nowak98JTB}. Alternatively, if players
can choose not to play the public goods games, such players called
loners beat ALLD,
ALLD beats ALLC, and ALLC beats loners.  The elicited RSP dynamics save
altruistic ALLC players, which would not survive without loners 
\cite{Hauert02sci,Hauert02jtb,Semmann03}. In the iterated Prisoner's
Dilemma, a mixture of ALLC, ALLD, and
tit-for-tat (TFT) constitutes RSP dynamics so that altruism survives
under small mutation rates \cite{Imhof05}. Finally, a population of
different types of strategies, each of which is
based only on the last actions of players, shows periodic and
chaotic oscillations of population densities associated with
cyclic competition \cite{Nowak89jamc,Nowak89jtb,Nowak93pnas,Brandt06jtb}.

As illustrated by these examples, an important consequence
of cyclically competing evolutionary dynamics is oscillations of
population densities. Each strategy or phenotype alternately becomes
the majority until it is devoured by its enemy. Oscillations are
usually self-destructive in small populations
because the finite-size effect usually drives the evolutionary
dynamics to eventual dominance of one of the competing phenotypes
\cite{Taylor78,Nowak98JTB,Reichenbach06pre}.  If the model represents
a ecosystem, dominance of one species implies a loss of
biodiversity. In social dilemma games, a situation in which
one strategy wins
cyclic competition often implies a loss of altruism;
in many models, an altruistic strategy is relatively weak
compared to others and beats another strategy only
barely \cite{Nowak98JTB,Hauert02jtb,Imhof05}.

If oscillations are stabilized or suppressed, cyclically competing
strategies will coexist in a population.
Oscillations can be suppressed by introduction of spatial structure such
as the square lattice
\cite{Hassell91,Tainaka93,Durrett98TPB,Frean01}.
In line with this, oscillation amplitudes grow as 
the underlying population structure
transits from 
the regular lattice to the
well-mixed population by increasing the number of
long-range edges
\cite{Szabo04jpa,Ying07jpa} (see \cite{Szabo07physr} for more references).
Alternatively,
heterogeneous contact rates of individuals suppress oscillations.
In this scenario, hubs, namely, individuals 
with many neighbors, are more susceptible to
invasion by their enemies,
because hubs have many enemies in the neighborhood
than non-hubs on average.
Then hubs complete a state-transition 
cycle in shorter periods. Mixture of individuals
with different oscillation periods abolishes the oscillation of the
entire population density \cite{MasudaRSP}.

An implicit assumption made by these and many other studies of cyclic
competition \cite{Szabo07physr} is that a focal individual is invaded
by each enemy neighbor independently at a constant rate.  For example,
if a focal individual is in state R and it has 10 neighbors, the state
of the focal individual is replaced by state P, which governs R, at a
rate proportional to the number of P among the 10 neighbors. In other
words, each P individual preys on each R neighbor independently at a
constant rate. The contributions of different P individuals add
linearly.  In evolutionary games, however, the effect of each neighbor
adds linearly in the sense that
the aggregated payoff of the focal player over is the linear sum
of the payoff obtained by playing with each neighbor.
Then natural selection operates on the
aggregated payoff in a nonlinear manner.  Here how one player elicits
state transition of another player is not straightforward.

In addition, suppression of oscillations by the heterogeneity
mechanism requires that the number of contacts, or
neighbors, depends on individuals \cite{MasudaRSP}.  In the context
of evolutionary games, heterogeneous contact numbers also pave the way
to increased
altruism \cite{Santos05prl,Santos06pnas}. In this situation,
enhanced cooperation stems
from players with many contacts, who earn more by
participating in the game more often. However, how to compare
aggregated payoffs of players with different numbers of participation
is an unresolved issue \cite{MasudaPDSF}.

In this work, we explore how oscillations in RSP dynamics can be
stabilized or suppressed in evolutionary games.  To avoid the problem
of the heterogeneous number of contact per player, we consider a
well-mixed population and assume that the adaptation rate, but not the
contact number, depends on players.  Some players are quicker to copy
successful strategies than others.  Such heterogeneous adaptation
rates are known to enhance cooperation in the evolutionary Prisoner's
Dilemma on the regular lattices and small-world networks
\cite{Kim02pre,Szolnoki07epl}.  By extending the replicator dynamics,
we show that heterogeneous adaptation rates suppress oscillations of
population densities in well-mixed populations.
We validate our theory by numerical simulations
of the RSP games and the public goods game with loners
\cite{Hauert02sci,Hauert02jtb}.

\section{Model}

\subsection{Replicator dynamics}

We begin with the standard replicator dynamics.
Suppose that there are $m$ possible phenotypes, which we call strategies.
The proportion of individuals, which we call players,
with strategy $i$ in the population of $n$ players is denoted by
$x_i$ ($1\le i\le m$), where
$\sum_i x_i = 1$ and $0\le x_i\le 1$ ($1\le i\le m$) are always satisfied.
The fitness of strategy $i$ is denoted by $f_i$.
The standard replicator dynamics are represented by
\begin{eqnarray}
\frac{dx_i}{dt} &=& x_i \left( f_i  - \sum_{j=1}^m
f_j x_j \right)\nonumber\\
&=& x_i \sum_j \left( f_i - f_j \right) x_j,
\label{eq:repl}
\end{eqnarray}
where we suppress the range of $j$ when no confusion arises.
Players with strategy $i$ increase in number at a rate proportional to
the excess of $f_i$ relative to the average fitness $\sum_j f_j x_j$.
Put another way, 
a player with strategy $j$ switches its strategy to
strategy $i$ of a randomly picked opponent 
at a rate proportional to the fitness difference $f_i-f_j$, if
$f_i-f_j\ge 0$. If $f_i-f_j<0$, players with strategy 
$i$ may switch to strategy $j$.
The players are interpreted to form a well-mixed population.

We rewrite \EQ(\ref{eq:repl}) as
\begin{equation}
\frac{dx_i}{dt} = x_i 
\sum_j G(f_i-f_j)x_j - x_i \sum_j G(f_j-f_i)x_j,
\label{eq:G}
\end{equation}
where
\begin{equation}
G(x) = \left\{ \begin{array}{ll}
x, & (x\ge 0)\\
0. & (x < 0)
\end{array}\right.
\label{eq:G-piecewise}
\end{equation}
The first term of \EQ(\ref{eq:G}) represents the inflow of the population density
to strategy $i$ and the second term represents the outflow from $i$ to
other strategies. Note that
\begin{equation}
G(x)-G(-x)=x
\label{eq:property-G}
\end{equation}
is satisfied.

\subsection{Extended replicator dynamics}\label{sub:extended}

We extend the standard replicator dynamics by considering a
heterogeneous population of players who differ in the adaptation
rate.  For simplicity, we assume only two adaptation rates $\beta_x$
and $\beta_y$.  The corresponding proportions of each strategy in two
different subpopulations are specified by $x_i$ and $y_i$ ($1\le i\le
m$), where $\sum_i x_i = \sum_i y_i = 1$, and $0\le x_i, y_i\le 1$
($1\le i\le m$).  In a homogeneous population, the adaptation rate,
which is common to all the players, can be ignored by rescaling the
time, as is evident in \EQ(\ref{eq:repl}).  With two adaptation rates,
the ratio $\beta_x/\beta_y$ is an essential parameter that may alter
dynamical consequences. Note that the interaction
rate is common to all the players, which contrasts with 
Taylor and Nowak (2006).
%
%\cite{Taylor06}.

If one strategy is more profitable than another strategy irrespective
of the values of $x_i$ and $y_i$ ($1\le i\le m$), 
players that adapt more rapidly
enjoy more benefits. For example, in the Prisoner's Dilemma with ALLC
and ALLD strategies only, everybody in a well-mixed population will
choose ALLD.  Rapidly adapting ALLC players switch to ALLD in early
stages to transiently exploit slowly adapting ALLC players. However,
under cyclic competition, it may not be profitable to myopically
follow apparently successful others.  In a long term, players may be
better off by adapting slowly.

We assume that players differ only in adaptation rates and 
that they are well-mixed.
Therefore, the total payoff of a player with strategy $i$ is 
equal to $f_i$
regardless of the adaptation rate.
We represent the extended replicator dynamics as follows:
\begin{eqnarray}
\frac{1}{\beta_x}\frac{dx_i}{dt} &=& x_i 
\left(f_i - \sum_j f_j x_j \right)
+ y_i \sum_j G(f_i-f_j)x_j
- x_i \sum_j G(f_j-f_i)y_j
\label{eq:repl-extended1}\\
\frac{1}{\beta_y}\frac{dy_i}{dt} &=& y_i
\left(f_i - \sum_j f_j y_j\right)
+ x_i \sum_j G(f_i-f_j)y_j
- y_i \sum_j G(f_j-f_i)x_j.
\label{eq:repl-extended2}
\end{eqnarray}
In \EQ(\ref{eq:repl-extended1}),
the first term in the right-hand side represents the competition
within players that have adaptation rate $\beta_x$. This part is identical to
the standard replicator dynamics.
The second term represents the transition
from strategy $j$ to strategy $i$ as a result of imitating
a player with the different adaptation rate, that is, $\beta_y$.
Players with strategy $j$ and adaptation
rate $\beta_x$, whose density is $x_j$, switch to strategy $i$ when they
meet a player with strategy $i$ and adaptation rate $\beta_y$. The transition
rate is proportional to $G(f_i-f_j)$.
Note that the transition from $j$ to $i$ such that
players with strategy $i$ and adaptation rate $\beta_x$ are imitated
is taken care of by the first term.
The third term represents the rate at which
players with strategy $i$ and adaptation rate $\beta_x$ switch to
strategy $j$ when they meet a player with strategy $j$ and adaptation
rate $\beta_y$. Equation (\ref{eq:repl-extended2}) can be
interpreted similarly.
Equations~(\ref{eq:repl-extended1})
and (\ref{eq:repl-extended2}) guarantee
$\sum_i dx_i/dt = 0$ and $\sum_i dy_i/dt = 0$, so that
the number of players in each subpopulation is conserved.

Cyclic competition often yields coexistence of all the relevant strategies.
Interested in this phenomenon,
we focus on interior equilibria, which satisfy
$x_i>0$ ($1\le \forall i\le m$).  
In the standard replicator dynamics shown in
\EQ(\ref{eq:repl}), suppose that
the total payoff for each strategy $f_i$ is linear in
$x_1$, $x_2$, $\ldots$, $x_m$, as is the case when the payoff is
defined by a payoff matrix. Then the interior equilibrium is unique and
determined from $f_1$ $=$ $f_2$ $=$ $\ldots$ $=$ $f_m$ and $\sum_i x_i = 1$
\cite{Zeeman80,Hofbauerbook}.  For the extended replicator dynamics, the
interior equilibrium is not unique even
when $f_i$ is linear in the density of each strategy $x_1+y_1$, 
$x_2+y_2$, $\ldots$, and $x_m+y_m$. Indeed, any
configuration that satisfies $f_1$ $=$ $f_2$ $=$ $\ldots$ $=$ $f_m$, $\sum_i
x_i = 1$, and $\sum_i y_i = 1$ leads to $dx_i/dt = dy_i/dt = 0$ in
\EQS(\ref{eq:repl-extended1}) and (\ref{eq:repl-extended2}) for $G$ given in
\EQ(\ref{eq:G-piecewise}). Because there are $2m$ unknown variables and $m+1$
equations, the equilibrium is underdetermined.
Only the proportion of each strategy with 
different adaptation rates pooled, namely, $x_i+y_i$, is determined by the
equilibrium condition.

To avoid this ambiguity, we introduce a realistic assumption for
$G$. Because of bounded cognitive ability and stochastic environments,
players may mimic strategies of unsuccessful others with small
probabilities. This factor can be implemented
by allowing $G(x)>0$ for $x<0$
\cite{Blume93,Nowak04nat,Ohtsuki06nat,Traulsen06pre}.
More specifically, we assume that $G$ is any differentiable function
that satisfies $G^{\prime}(x)>0$, $\lim_{x\to-\infty}G(x)=0$, and
\EQ(\ref{eq:property-G}).
This assumption
implies that $\lim_{x\to\infty}G(x)/x= \lim_{x\to\infty}(x+G(-x))/x=
1$ and $G^{\prime}(0)=1/2$. 

Then, it is straightforward to see that
the interior equilibrium in the case of
the linear payoff is uniquely
given by $f_1$ $=$ $f_2$ $=$ $\ldots$ $=$ $f_m$, $\sum_i
x_i = 1$, $\sum_i y_i = 1$, and 
$x_i=y_i$ ($1\le i\le m$).
The uniqueness can be shown as follows.
If $f_1$ $=$ $f_2$ $=$ $\ldots$ $=$ $f_m$ does not hold
at an interior equilibrium,
there exists an $i_0$
such that $f_{i_0}>\sum_j f_j x_j$
and $f_{i_0}>\sum_j f_j y_j$ at this point.
For this $i_0$, summation of
\EQS(\ref{eq:repl-extended1}) and
(\ref{eq:repl-extended2}) yields
\begin{equation}
0 = \frac{1}{\beta_x}\frac{dx_{i_0}}{dt} + 
\frac{1}{\beta_y}\frac{dy_{i_0}}{dt}
= x_{i_0}\left(f_{i_0}-\sum_j f_jx_j\right)
+ y_{i_0}\left(f_{i_0}-\sum_j f_jy_j\right)>0,
\end{equation}
which is a contradiction. Therefore, 
$f_1$ $=$ $f_2$ $=$ $\ldots$ $=$ $f_m$ holds.
Then the first terms in the right-hand sides of  
\EQS(\ref{eq:repl-extended1}) and
(\ref{eq:repl-extended2}) vanish, and we obtain $x_i=y_i$ ($1\le i\le m$).

\section{Analysis of the RSP games}\label{sec:rsp-analytical}

We consider the symmetric RSP game whose
fitness $f_i$ is specified via the following payoff matrix:
\begin{equation}
A=\left(\begin{array}{ccc}
0&-a_2&b_3\\
b_1&0&-a_3\\
-a_1&b_2&0
\end{array}\right)
\label{eq:matrix-nonzerosum}
\end{equation} 
\cite{Zeeman80,Hofbauerbook}. Each player can take one of the three
strategies ($m=3$), and the row player and the column player
correspond to the focal player and the opponent, respectively.

The unique interior equilibrium of the extended replicator dynamics
is given by
\begin{equation}
\left(\begin{array}{c}
x_1^*\\x_2^*\\x_3^*
\end{array}\right)
= \left(\begin{array}{c}
y_1^*\\y_2^*\\y_3^*
\end{array}\right)
= \frac{1}{\Sigma}\left(\begin{array}{c}
a_2a_3+a_3b_2+b_2b_3\\
a_1a_3+a_1b_3+b_1b_3\\
a_1a_2+a_2b_1+b_1b_2
\end{array}\right),
\end{equation}
where $\Sigma$ is the normalization constant.  In the standard
replicator dynamics, this equilibrium is asymptotically stable when
$a_1a_2a_3 < b_1b_2b_3$ \cite{Zeeman80,Hofbauerbook}.  If
$a_1a_2a_3>b_1b_2b_3$, the equilibrium is unstable, and in a finite
population, the population density oscillates with ever increasing
amplitudes until the population consists entirely of a single
strategy. The relation $a_1a_2a_3=b_1b_2b_3$ stipulates a borderline
situation in which the interior equilibrium is neutrally stable.

By a one-to-one mapping of the population density vector,
the payoff matrix $A$ can be transformed to the following form
\cite{Zeeman80,Hofbauerbook}:
\begin{equation}
A^{\prime}=\left(\begin{array}{ccc}
0&-a_2&a_1+\epsilon\\
a_2+\epsilon&0&-a_3\\
-a_1&a_3+\epsilon&0
\end{array}\right)
\label{eq:matrix-nonzerosum2}
\end{equation} 
Now, the asymptotic stability of the interior equilibrium is
equivalent to $\epsilon>0$.

To show that heterogeneous adaptation rates
stabilize the interior equilibrium, we even simplify
\EQ(\ref{eq:matrix-nonzerosum2}) by setting $a_1=a_2=a_3=1$ 
\cite{Zeeman80}:
\begin{equation}
A^{\prime\prime} = \left(\begin{array}{ccc}
0&-1&1+\epsilon\\
1+\epsilon&0&-1\\
-1&1+\epsilon&0
\end{array}\right)
\label{eq:matrix-nonzerosum3}
\end{equation} 
The interior equilibrium is now given by 
\begin{equation}
\left(\begin{array}{c}
x_1^*\\x_2^*\\x_3^*
\end{array}\right)
= \left(\begin{array}{c}
y_1^*\\y_2^*\\y_3^*
\end{array}\right)
= \frac{1}{3}\left(\begin{array}{c}
1\\ 1\\ 1
\end{array}\right),
\label{eq:*-simplified}
\end{equation}
which is asymptotically stable if and only if
$\epsilon > 0$. When $\epsilon=0$, the equilibrium is neutrally stable,
with $A^{\prime\prime}$ defining a zerosum game. The standard replicator
dynamics for $\epsilon=0$ becomes the most famous RSP dynamics:
\begin{eqnarray}
\frac{dx_1}{dt} &=& x_1(x_3-x_2),\\
\frac{dx_2}{dt} &=& x_2(x_1-x_3),\\
\frac{dx_3}{dt} &=& x_3(x_2-x_1).
\label{eq:rsp-standard}
\end{eqnarray}

For $A^{\prime\prime}$ given in \EQ(\ref{eq:matrix-nonzerosum3}), we obtain
\begin{eqnarray}
f_1 &=& (1+\epsilon)(x_3+y_3)-(x_2+y_2),\\
f_2 &=& (1+\epsilon)(x_1+y_1)-(x_3+y_3),\\
f_3 &=& (1+\epsilon)(x_2+y_2)-(x_1+y_1).
\end{eqnarray}
We linearize the extended replicator dynamics
around the interior equilibrium by setting
$x_i = x_i^*+\Delta x_i$, $y_i = y_i^*+\Delta y_i$ ($1\le i\le 3$),
where $\Delta x_i$ and $\Delta y_i$ are small.
Using $\Delta x_1+\Delta x_2+\Delta x_3 = \Delta y_1
+\Delta y_2 + \Delta y_3 = 0$ and
$G^{\prime}(0) = 1/2$, we obtain
\begin{eqnarray} 
&&\frac{1}{\beta_x}\frac{d\Delta x_1}{dt}\nonumber\\
&=& \left(\frac{1}{3}+\Delta x_1\right)
\left[\left(1+\epsilon\right)
\left(\Delta x_3 + \Delta y_3\right)
- \left(\Delta x_2 + \Delta y_2\right)\right.\nonumber\\
&-& \left.
\epsilon\left(\Delta x_1 + \Delta x_2 + \Delta x_3\right)
-\frac{\epsilon}{3}\left(\Delta y_1 + \Delta y_2 + \Delta y_3\right)
\right]\nonumber\\
&+& \left(\frac{1}{3}+\Delta y_1\right)
\sum_j\left[G\left(0\right)+G^{\prime}\left(0\right)
\left(\Delta f_1 - \Delta f_j\right)\right]
\left(\frac{1}{3}+\Delta x_j\right)\nonumber\\
&-& \left(\frac{1}{3}+\Delta x_1\right)
\sum_j\left[G\left(0\right)+G^{\prime}\left(0\right)
\left(\Delta f_j - \Delta f_1\right)\right]
\left(\frac{1}{3}+\Delta y_j\right)\nonumber\\
&=& \left[-\frac{2(1+\epsilon)}{3}-G\left(0\right)\right]\Delta x_1
-\frac{2(2+\epsilon)}{3}\Delta x_2\nonumber\\
& & + \left[-\frac{2(1+\epsilon)}{3}+G\left(0\right)\right]\Delta y_1
-\frac{2(2+\epsilon)}{3}\Delta x_2,
\end{eqnarray}
where $\Delta f_i$ is perturbation of the steady-state $f_i$
that originates from $\Delta x_1$, $\Delta x_2$, and $\Delta x_3$.
By doing similar calculations for
$\Delta x_2$, $\Delta y_1$, and $\Delta y_2$,
we obtain the linear dynamics around the equilibrium
given by \EQ(\ref{eq:*-simplified}) as follows:
\begin{equation}
\frac{d}{dt}\left(\begin{array}{c}
\Delta x_1\\ \Delta x_2\\ \Delta y_1\\ \Delta y_2
\end{array}\right)
=
\left(\begin{array}{cccc}
-(1+\epsilon)B_x-C_x& -(2+\epsilon)B_x & -(1+\epsilon)B_x+C_x & -(2+\epsilon)B_x\\
(2+\epsilon)B_x & B_x-C_x & (2+\epsilon)B_x
& B_x+C_x\\
-(1+\epsilon)B_y+C_y & -(2+\epsilon)B_y & 
-(1+\epsilon)B_y-C_y & -(2+\epsilon)B_y\\
(2+\epsilon)B_y &B_y+C_y &
(2+\epsilon)B_y & B_y-C_y
\end{array}\right)
\left(\begin{array}{c}
\Delta x_1\\ \Delta x_2\\ \Delta y_1\\ \Delta y_2
\end{array}\right),
\end{equation}
where $B_x = 2\beta_x/3$, $B_y = 2\beta_y/3$,
$C_x=\beta_xG(0)$, and $C_y=\beta_yG(0)$.
Noting that $B_xC_y = B_yC_x$, we obtain
the following eigenvalue equation:
\begin{eqnarray}
&&\lambda^4 + \left[2\left(C_x+C_y\right)
+\epsilon\left(B_x+B_y\right)\right]\lambda^3\nonumber\\
&+& \left[\left(3+3\epsilon+\epsilon^2\right)\left(B_x^2+B_y^2\right)
+\left(C_x+C_y\right)^2+\left(6+6\epsilon+\epsilon^2\right)B_xB_y\right.
\nonumber\\
&+&\left.\epsilon \left(B_xC_x + B_yC_y + 3B_xC_y + 3B_yC_x\right)
\right]\lambda^2\nonumber\\
&+& \left[ 6\left(2+\epsilon\right)^2 B_xB_y\left(C_x+C_y\right)
+ 4\epsilon\left(B_x+B_y\right)C_xC_y
 \right]\lambda + 12(2+\epsilon)^2 B_xB_yC_xC_y = 0.
\label{eq:eigen}
\end{eqnarray}

The interior
equilibrium is stable
if all the solutions of \EQ(\ref{eq:eigen}) have negative real parts.
The Routh-Hurwitz criteria dictate that all the eigenvalues of 
\EQ(\ref{eq:eigen}) are negative if and only if
the four principal minors of a matrix calculated from the coefficients of 
\EQ(\ref{eq:eigen}) are all positive. To the first order of $\epsilon$,
the principal minors are calculated as:
\begin{eqnarray}
|H_1| &=& 2(C_x+C_y)+\epsilon(B_x+B_y),
\label{eq:routh-H1}
\\
|H_2| &=& 6(1+\epsilon)\left(B_x-B_y\right)^2(C_x+C_y)
+2\left(C_x+C_y\right)^3 +3\epsilon(B_x+B_y)^3\nonumber\\
&+& \epsilon(B_x+B_y)(3C_x^2+10C_xC_y+3C_y^2)
\label{eq:routh-H2}\\
|H_3| &=& 144(1+2\epsilon)(B_x-B_y)^2B_xB_y(C_x+C_y)^2
+48(1+\epsilon)B_xB_y(C_x+C_y)^2(C_x-C_y)^2\nonumber\\
&+& 72\epsilon(B_x+B_y)^3B_xB_y(C_x+C_y)
+24\epsilon(B_x+B_y)B_xB_y(C_x+C_y)(2C_x^2+3C_xC_y+2C_y)^2\nonumber\\
&+& 24\epsilon(B_x+B_y)(B_x-B_y)^2(C_x+C_y)C_xC_y
+8\epsilon (B_x+B_y)(C_x+C_y)^3C_xC_y,
\label{eq:routh-H3}\\
|H_4| &=& 48(1+\epsilon)B_xB_yC_xC_y|H_3|
\label{eq:routh-H4}
\end{eqnarray}
These relations hold whenever $\epsilon>0$, that is, when the interior
equilibrium is stable even without heterogeneous adaptation
rates. When $\epsilon<0$, the interior equilibrium is unstable
with homogeneous adaptation rates.
In accordance, $|H_3|<0$ holds when
$B_x=B_y>0$, $C_x=C_y>0$, and $\epsilon<0$.
To make $|H_3|$ positive for $\epsilon<0$, the heterogeneity quantified by
$(B_x-B_y)^2$ and $(C_x-C_y)^2$,
must be large enough to compensate the negative
contribution of the last four $\epsilon$ terms of
\EQ(\ref{eq:routh-H3}) to $|H_3|$.

Although \EQS(\ref{eq:routh-H1})--(\ref{eq:routh-H4}) are valid for
infinitesimally small $|\epsilon|$, they are exact when $\epsilon=0$.
In this case, the standard replicator dynamics have a neutrally stable
interior equilibrium \cite{Zeeman80,Hofbauerbook}. Actually, $|H_3|=0$
holds when $B_x=B_y>0$, $C_x=C_y>0$, and $\epsilon=0$.
Equations~(\ref{eq:routh-H1})--(\ref{eq:routh-H4}) imply that the
asymptotic stability is equivalent to $B_x\neq B_y$ and $C_x\neq C_y$,
or equivalently, $\beta_x\neq \beta_y$, in addition to $\beta_x>0$ and
$\beta_y>0$. These analytical results indicate that heterogeneous
adaptation rates stabilize the interior equilibrium.

\section{Numerical results}\label{sec:numerical}

In this section, we perform
individual-based numerical 
simulations in which players are involved in
evolutionary games with
cyclic competition. As a rule of thumb,
when the population size is of the order of 100 or
smaller, the finite-size effect tends to make amplitudes of 
oscillatory population densities explode.
We show that, even in this case,
introduction of 
heterogeneous adaptation rates prohibits the monopoly by a single
strategy, which would follow unrestricted growths
of oscillation amplitudes.
As a result, different competing strategies can coexist.
In the following numerical simulations, we assume
$n=200$ players. 

\subsection{RSP games}\label{sub:rsp-numerical}

Consider that players are involved in the RSP game whose payoff
matrix is given by \EQ(\ref{eq:matrix-nonzerosum3}).  Each player
takes one of the three strategies.  Initially, players select each
strategy randomly and independently with probability $1/3$.  Any pair
of players is engaged in the game with probability 0.5 so that each
player is matched with $0.5(n-1)=99.5$ other players on average.

In accordance with the replicator dynamics, players with larger
accumulated payoffs can disseminate their strategies more
successfully.  We pick 40 pairs of players randomly from the
population and denote one such pair by $i$ and $j$.  Only player $i$
is assumed to be subject to strategy update in this pairing.  Player
$i$ copies player $j$'s strategy with probability $\beta_i(f_j-f_i)$,
where $\beta_i$ is the adaptation rate of player $i$.  If $f_i>f_j$,
no imitation occurs. This update rule is equivalent to the replicator
dynamics with the piecewise linear $G$ given in
\EQ(\ref{eq:G-piecewise}). Although we required smooth $G$ for
formulating extended replicator dynamics, here we use the piecewise
linear $G$ for simplicity.
The average adaptation rate is set equal to
$\overline{\beta}=0.5/n=0.0025$,
and the adaptation rate of player $i$, or $\beta_i$,
is distributed according to the uniform density on
$\left[\overline{\beta}\left(1-\Delta_{\beta}\right),
\overline{\beta}\left(1+\Delta_{\beta}\right)\right]$,
independently for different players.
As $\Delta_{\beta}$ increases, the population
becomes heterogeneous.

\begin{figure}
\begin{center}
\includegraphics[height=2in,width=2in]{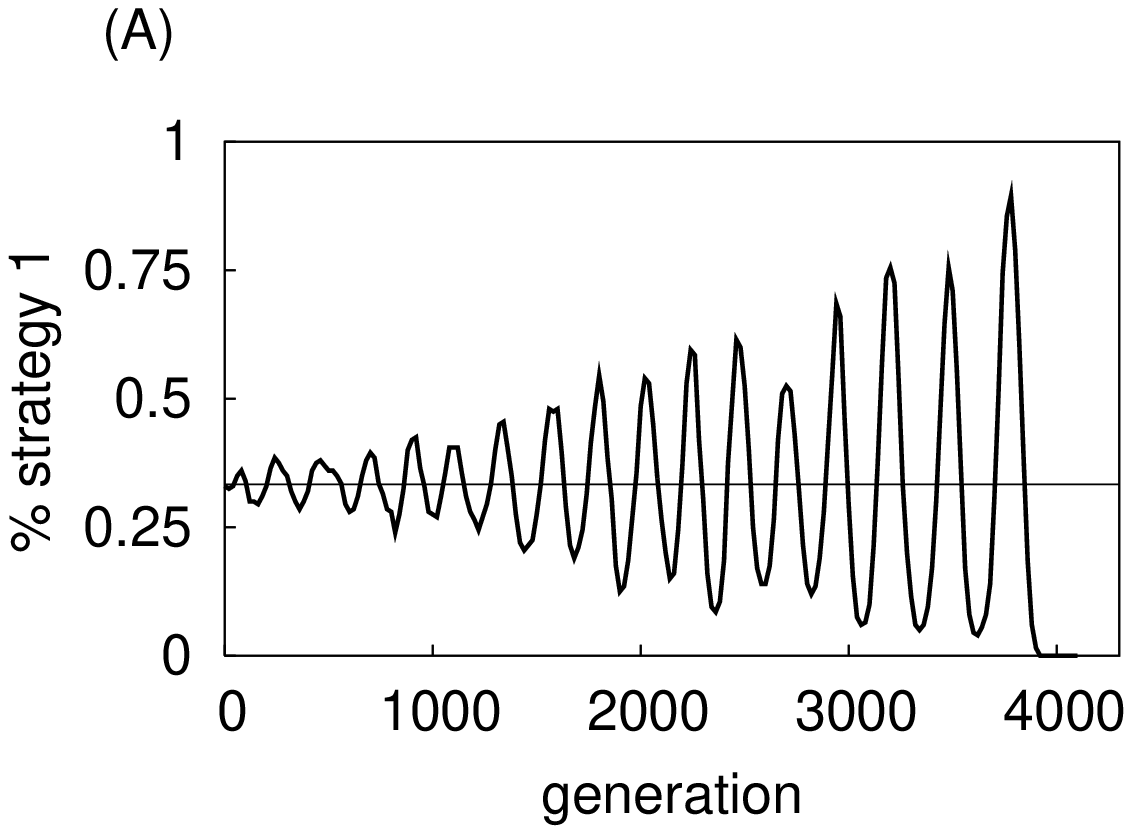}
\includegraphics[height=2in,width=2in]{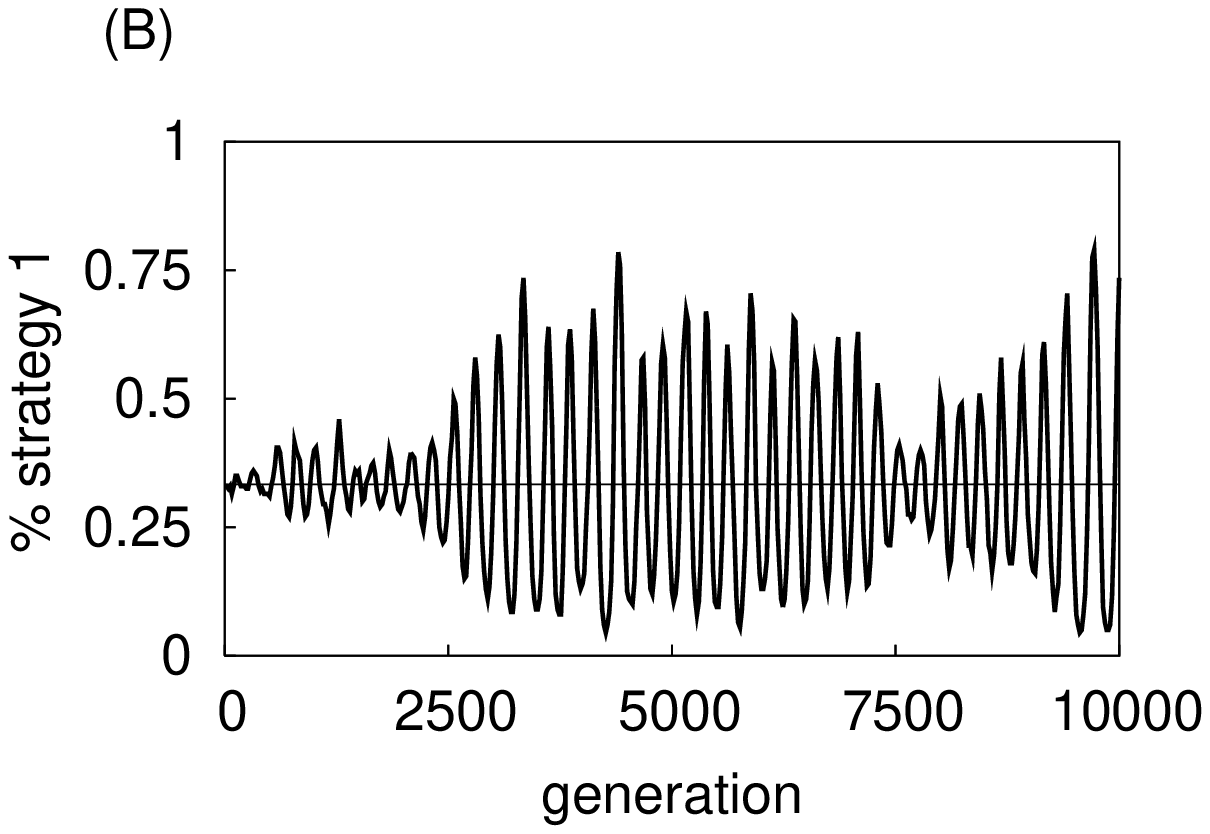}
\includegraphics[height=2in,width=2in]{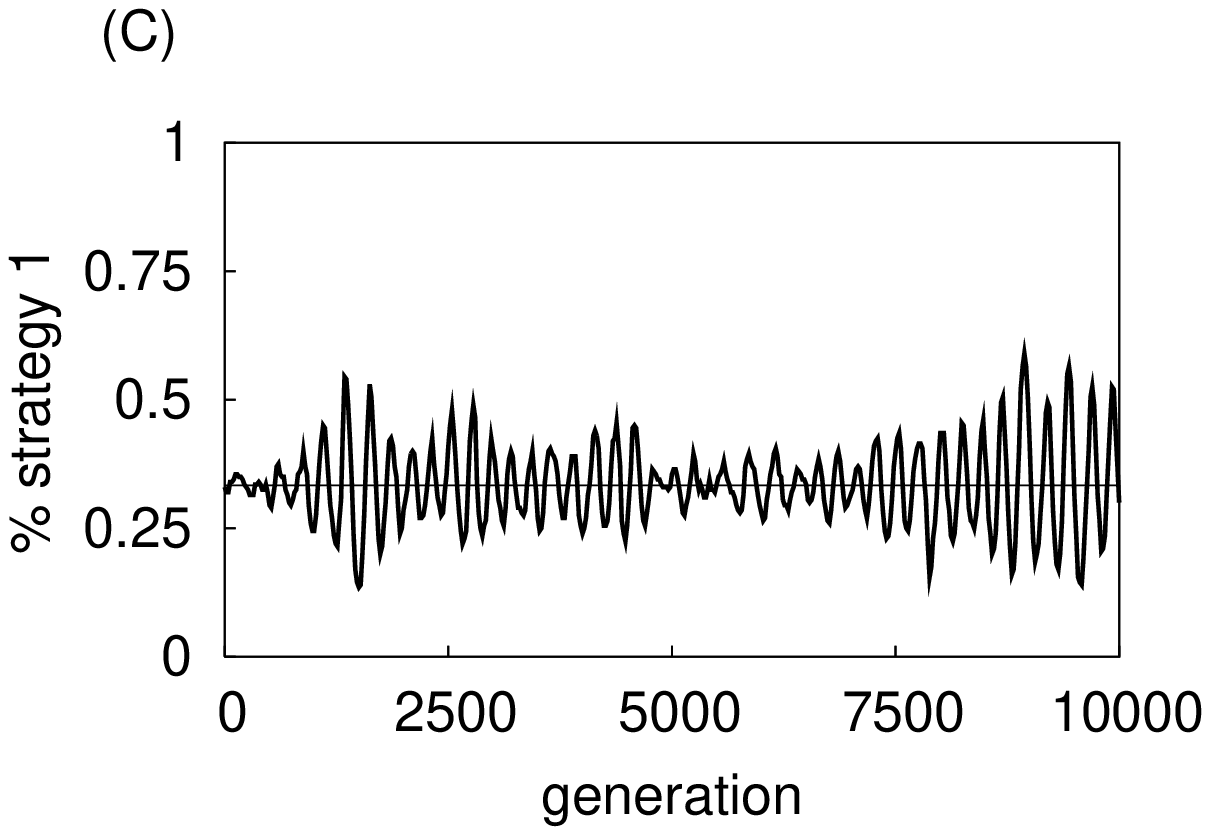}
\caption{Evolutionary simulations of the zerosum RSP game
in finite populations. We set $n=200$, $\epsilon=0$, and
$\overline{\beta}=0.0025$.
The dispersion of the adaptation rate is
(A) absent ($\Delta_{\beta}=0$), (B) intermediate
($\Delta_{\beta}=0.4$), and (C)
large ($\Delta_{\beta}=0.8$).
For clarity, the proportion of the players of only one 
of the three strategies is shown in each panel. The horizontal
lines indicate $1/3$.}
\label{fig:rsp-eps00}
\end{center}
\end{figure}

For $\epsilon=0$, the interior equilibrium is neutrally, but not
asymptotically, stable in the standard replicator dynamics.  The
results of individual-based simulations are shown in
\FIG\ref{fig:rsp-eps00} for three values of heterogeneity
$\Delta_{\beta}$. In \FIG\ref{fig:rsp-eps00}, only the proportions of
players with strategy 1 (first-row players in
\EQ(\ref{eq:matrix-nonzerosum3})) 
are shown for visibility. In reality, three
subpopulations, respectively, corresponding to strategies 1, 2, and 3
alternately prosper due to cyclic competition.

When $\Delta_{\beta}=0$ (\FIG\ref{fig:rsp-eps00}(A)), all the players
share the same adaptation rate ($\beta_i=\overline{\beta}$) as in the
standard replicator dynamics. Owing to the finite-size effect, the
oscillation grows in amplitude until one strategy dominates the
population after some oscillation cycles. The proportion of strategy 1
presented in the figure reaches 0 or 1 depending on initial conditions
and stochasticity in strategy updating.  As the adaptation rate
becomes heterogeneous, explosion of the oscillation amplitude tends to
be suppressed.  When $\Delta_{\beta}=0.4$
(\FIG\ref{fig:rsp-eps00}(B)), oscillations persist for much longer
time than when $\Delta_{\beta}=0$ (\FIG\ref{fig:rsp-eps00}(A)).  With
more heterogeneity, the oscillation persists even longer and the
oscillation amplitude becomes smaller, as shown in
\FIG\ref{fig:rsp-eps00}(C) for $\Delta_{\beta}=0.8$.

Precisely speaking, the theory predicts damping oscillations for
heterogeneous adaptation rates, but not reverberating ripples apparent
in \FIG\ref{fig:rsp-eps00}(B) and (C).  There are many factors that
could contribute to this discrepancy: the nonsmooth $G$, more than two
values of adaptation rates, stochasticity introduced by random
encountering and updating of randomly selected players, and small
$n$. Because the aim in the numerical simulations is to show
suppression of oscillations by heterogeneity, we do not explore this
subtle discrepancy.

\begin{figure}
\begin{center}
\includegraphics[height=2in,width=2in]{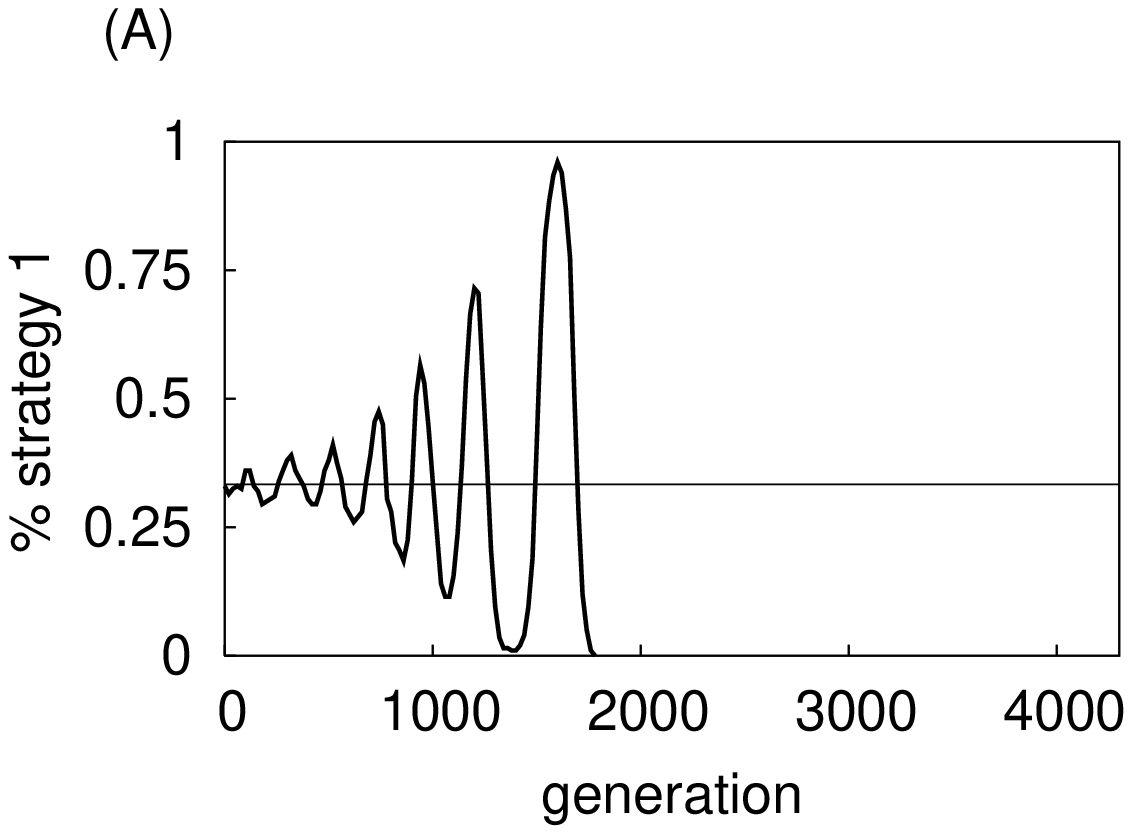}
\includegraphics[height=2in,width=2in]{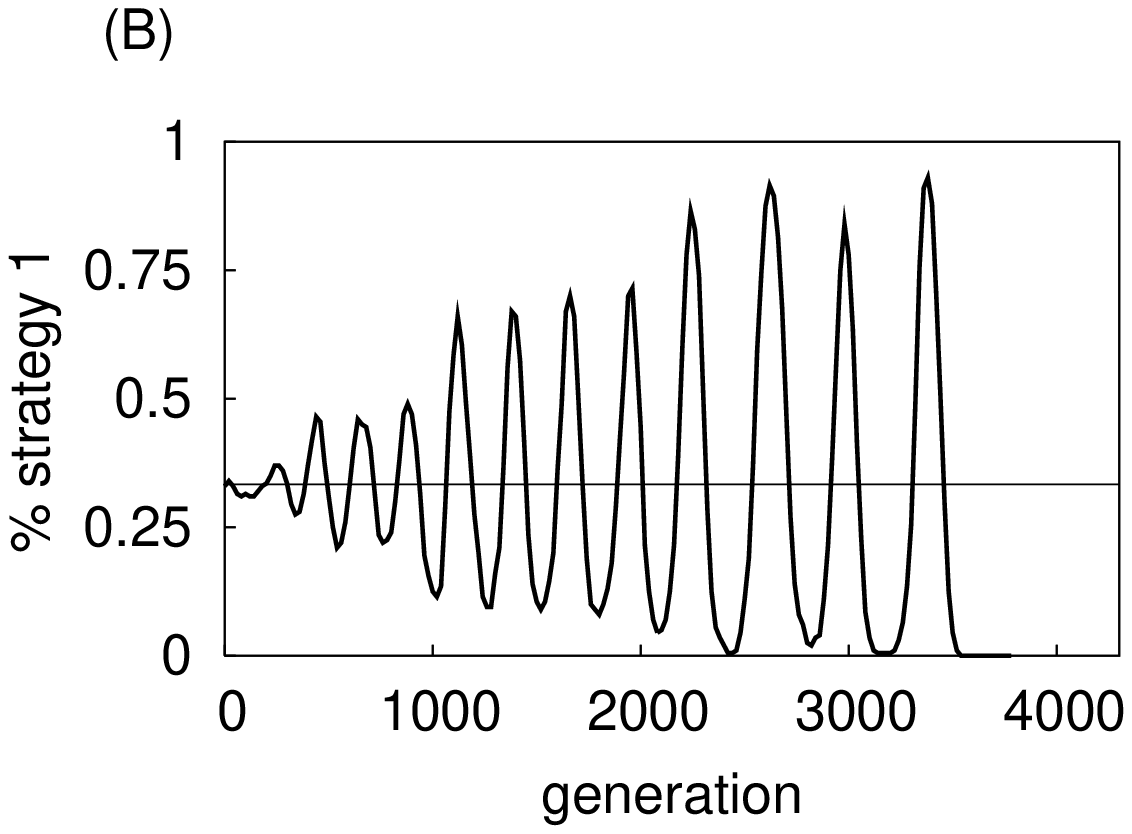}
\includegraphics[height=2in,width=2in]{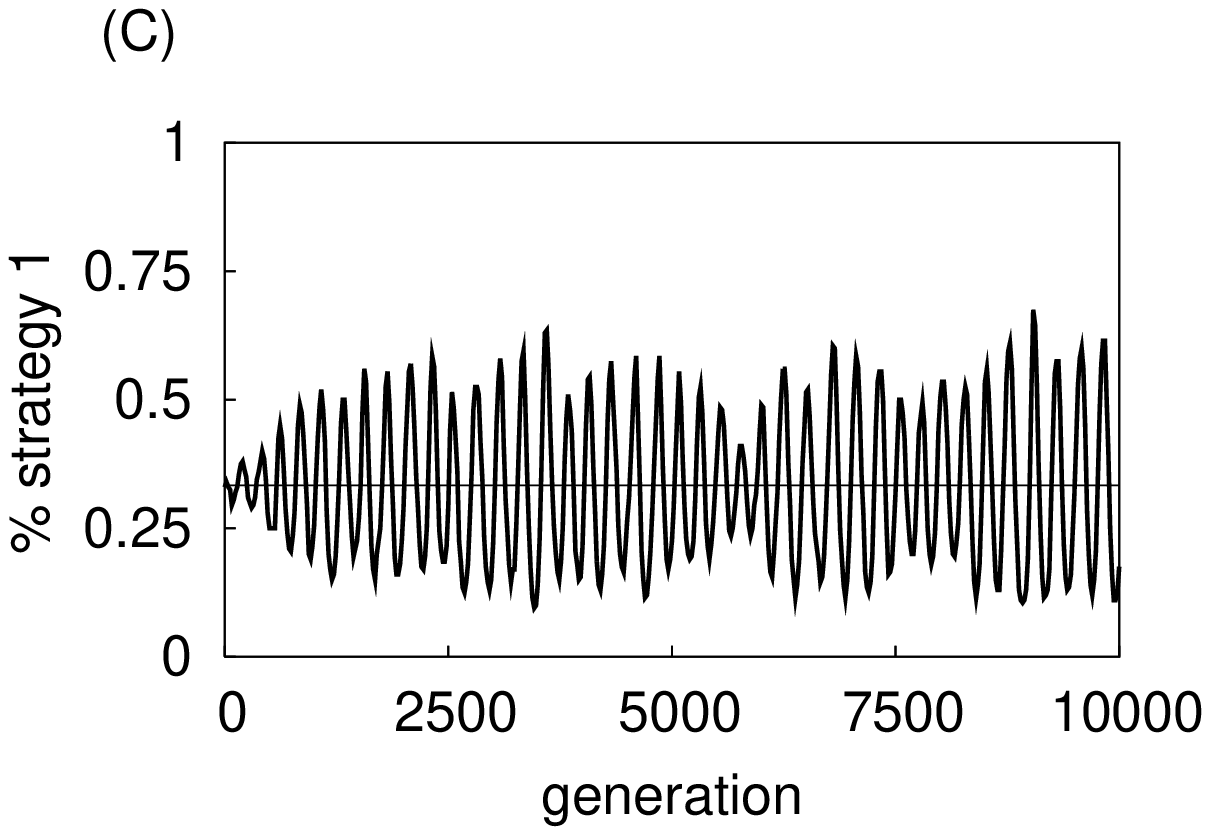}
\caption{Evolutionary simulations of the non-zerosum RSP game
in finite populations. We set $n=200$, $\epsilon=-0.1$,
$\overline{\beta}=0.0025$.
The dispersion of the adaptation rate $\Delta_{\beta}$ is equal to
(A) 0, (B) 0.4, and (C) 0.8.}
\label{fig:rsp-eps01}
\end{center}
\end{figure}

For $\epsilon=-0.1$, the interior equilibrium is unstable in the
standard replicator dynamics. In a homogeneous population, two of the
three strategies are eradicated in an early stage
(\FIG\ref{fig:rsp-eps01}(A); $\Delta_{\beta}=0$).  Even with some
heterogeneity in the adaptation rate, the oscillation ends up with
explosion (\FIG\ref{fig:rsp-eps01}(B); $\Delta_{\beta}=0.4$).  At this
level of heterogeneity, coexistence generally lasts longer than the
homogeneous case, whereas its duration depends pretty much on
numerical runs (results not shown). With stronger heterogeneity
(\FIG\ref{fig:rsp-eps01}(C); $\Delta_{\beta}=0.8$), three strategies
coexist stably with a high probability.

\subsection{Public goods game with voluntary participation}\label{sub:volu}

As another numerical example, we examine the public goods game, which
is a type of the multiperson Prisoner's Dilemma.  We assume that $N$
($<n$) randomly selected players form a group. In the standard public
goods game, each of $N$ players has an option to donate a unit cost or
to refrain from donation. The donated amount is multiplied by $r$ and
divided equally by the $N$ players. When $r>1$, donation is a
prosocial action because it increases the aggregated benefit of the
group by proportion $r-1$. 
However, a player always earns more by not donating,
which is a social dilemma.  In an evolutionary framework, the public
goods game is repeated with different random groups in one
generation. Then, the players with higher accumulated payoffs have
more chances to disseminate their strategies.  Without further
assumptions, altruistic donation is completely overridden by
defection.

Incorporation of another strategy called loner opens the way to
survival of cooperators \cite{Hauert02sci,Hauert02jtb}.  A loner does
not participate in the game, and it gains the side payoff
$\sigma>0$. The key assumption is that it is better not to play than
to be in a group of defectors. If there are $N-S$ loners in a group,
the other $S$ players play the public goods game. If there are $n_c$
cooperators among $S$ players given $S\ge 2$, the payoffs for a
cooperator, defector, and loner, are equal to $f_c = -1+r n_c/S$, $f_d
=r n_c/S$, and $f_l = \sigma$, respectively.  If $S=1$, the single
player constituting the group behaves like a loner because there is
nobody to play the game with.

With $\sigma<r-1$, it is better to stay in a group of cooperators than
not to participate.  Therefore, defectors dominate cooperators, as
before, and loners dominate defectors, and cooperators dominate
loners, which implies the RSP relation.  According to the meanfield
analysis with the standard replicator equations, the coexistence
equilibrium is neutrally stable \cite{Hauert02sci,Hauert02jtb}.  The
stable oscillation of the population density numerically shown for
$n=5000$ \cite{Hauert02sci} may collapse in small
populations. Oscillations can be stabilized by placing players on the
regular lattice \cite{Szabo02prl} or small-world networks with
sufficient spatial structure \cite{SzaboVukov04,Wu05pre} (see
\cite{Szabo07physr} for more references). Here we are concerned to an
alternative: the heterogeneity mechanism.

We perform numerical simulations with $n=200$,
$N=5$, $r=3$, and $\sigma=1$.
Initially, each player selects each 
strategy with probability $1/3$.
The generation payoff of player $i$ is determined as
the summation of the payoff after 2000 rounds of group formation.
As in the numerical simulations of the RSP game
(\SEC\ref{sub:rsp-numerical}), 
40 players are subject to strategy update at the end of each generation. 
The average adaptation rate is set equal to 
$\overline{\beta}=0.2N/n=0.005$.

\begin{figure}
\begin{center}
\includegraphics[height=2in,width=2in]{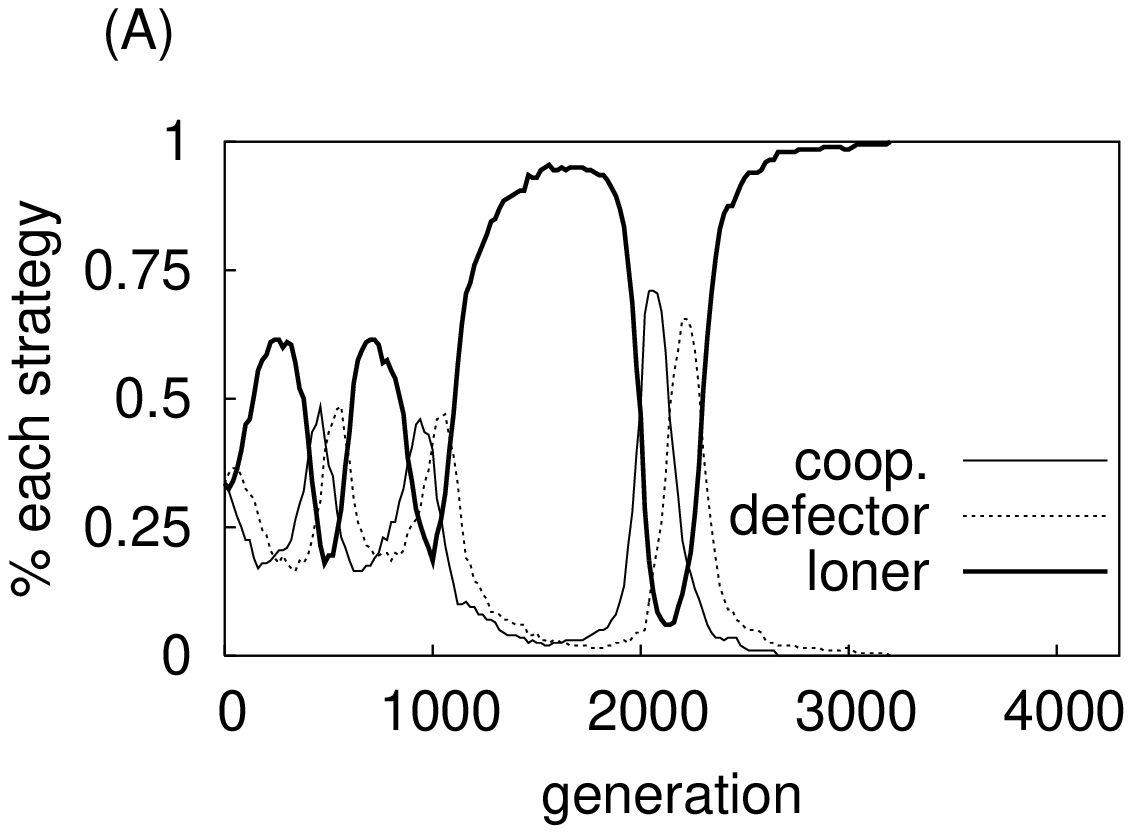}
\includegraphics[height=2in,width=2in]{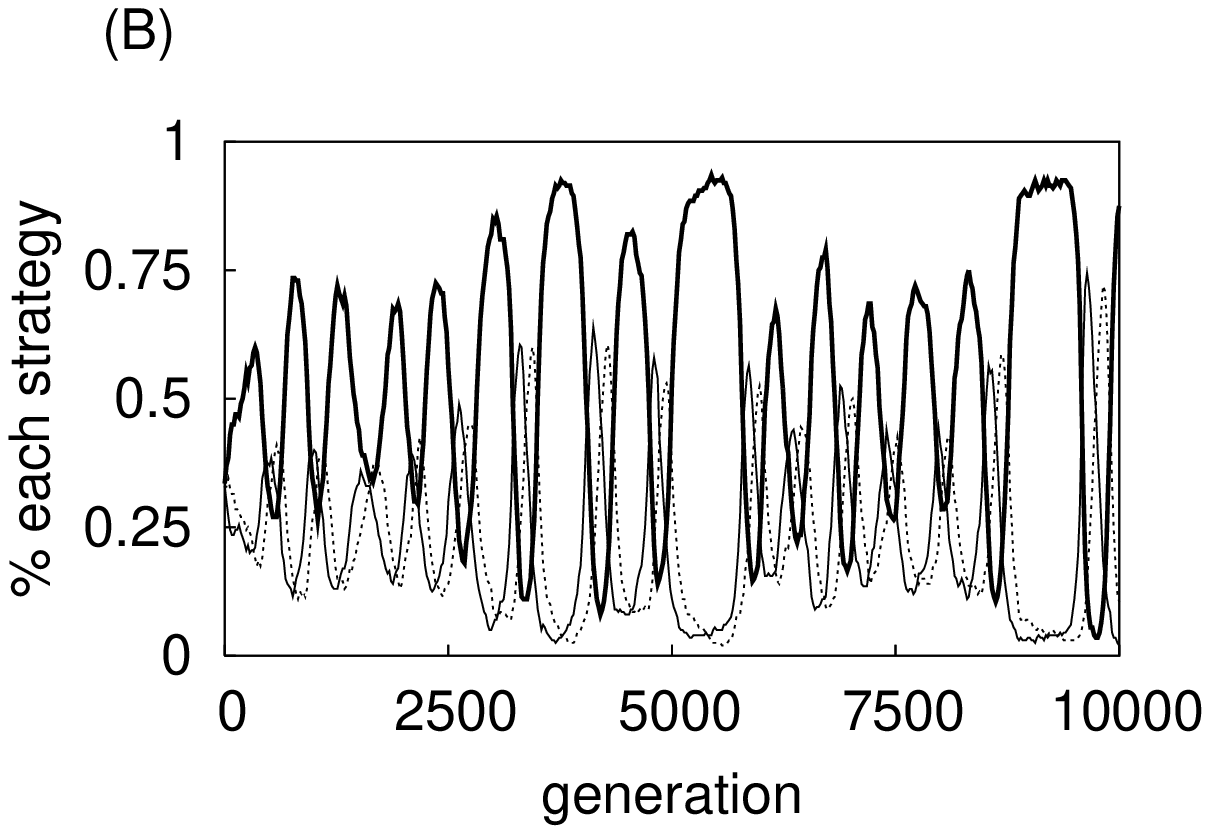}
\includegraphics[height=2in,width=2in]{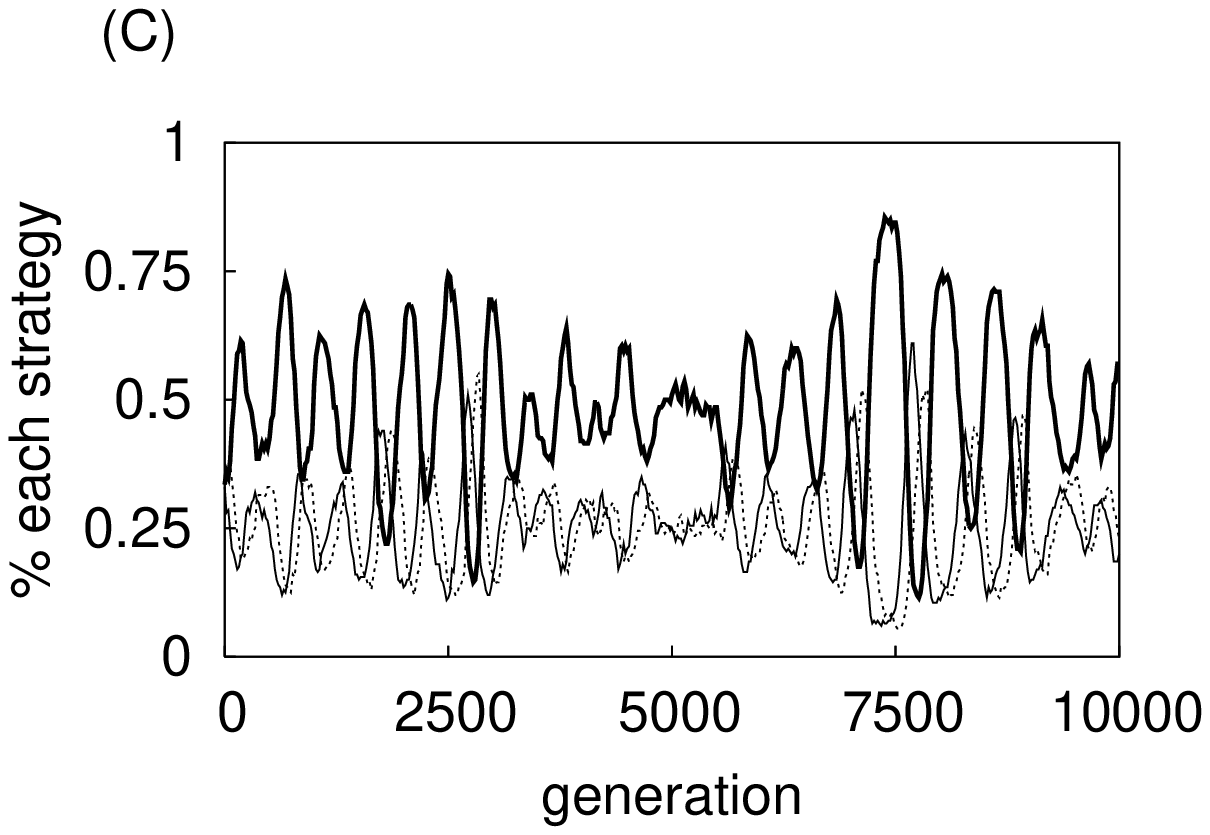}
\caption{Evolutionary simulations of the public goods game with
voluntary participation in finite populations. We set $n=200$, $N=5$,
$r=3$, $\sigma=1$, $\overline{\beta}=0.005$, and (A)
$\Delta_{\beta}=0$, (B) $\Delta_{\beta}=0.4$, and (C)
$\Delta_{\beta}=0.8$. The proportions of cooperators, defectors, and
loners, are shown by thin solid lines, thin dashed lines, and thick
solid lines, respectively.}
\label{fig:volu}
\end{center}
\end{figure}

The results are shown in \FIG\ref{fig:volu} for different degrees of
heterogeneity in the adaptation rate.  In each panel, cooperators
(thin solid lines) are devoured by defectors (thin dashed lines),
defectors are devoured by loners (thick solid lines), and loners are
devoured by cooperators. This completes one RSP cycle.  When the
adaptation rate is homogeneous as in the standard replicator dynamics,
the oscillation expands due to the finite-size effect.  Two of the
three strategies finally disappear (\FIG\ref{fig:volu}(A);
$\Delta_{\beta}=0$).  Unless $r$ is so large that loners are not
needed for the survival of cooperators, loners are relatively stronger
than cooperators and defectors without violating the RSP relationship
\cite{Hauert02sci,Hauert02jtb}.  Accordingly, the strategy that
dominates finite populations are usually loners, which contrasts to
the case of the RSP game.  When the adaptation rate is heterogeneous
(\FIG\ref{fig:volu}(B); $\Delta_{\beta}=0.4$), oscillations typically
last for longer time.  With a stronger degree of heterogeneity
(\FIG\ref{fig:volu}(C); $\Delta_{\beta}=0.8$), oscillations are more
stable.  In this way, heterogeneity promotes survival of cooperators
in small populations of players in the Prisoner's Dilemma.

\section{Discussion}

\subsection{Summary of the results}

We have analyzed evolutionary game dynamics of cyclically 
competing strategies (or phenotypes) with heterogeneous adaptation
rates. We have shown analytically and numerically that such
heterogeneity suppresses oscillations of the density of each strategy
and dominance of a particular strategy, which often occur in standard
evolutionary dynamics with cyclic competition.  In a game such that a
specific strategy is stable with a large attractive basin, players
with large adaptation rates shift to the lucrative strategy quickly. Then
they can transiently exploit conservative players with small
adaptation rates. A larger adaptation rate is
better in this situation.
However, in games with cyclic competition, an obvious
hierarchy of strategies is absent. Then conservative players
contribute to diversification of a population without really
sacrificing their own benefits.

The assumption that different players have different adaptation rates
seems realistic.  Therefore our results imply that cyclic competition
does not necessarily lead to apparent oscillations of population
densities or dominance of particular strategies, which many analytical
models predict. The other way round, an ordinary situation where three
or more types of strategists stably coexist does not exclude operation
of cyclic competition that is not weak.

Generally speaking, evolutionary dynamics in finite populations
deviate from theoretical predictions for infinite populations
\cite{Nowak04nat}.  In homogeneous populations, RSP dynamics that show
stable oscillations in infinite populations usually show unstable
oscillations in finite populations, so that all but one phenotypes
will disappear \cite{Reichenbach06pre}.  In the context of
evolutionary games, altruistic strategies that are relatively weak in
the RSP relation are expelled in finite populations, even if they can
survive in infinite populations.  The public goods game with voluntary
participation investigated in \SEC\ref{sub:volu} is such an example.
Heterogeneity saves rare strategies particularly in small populations
(up to the order of 100 players).

\subsection{Oscillations induced by mutation}

In addition to heterogeneous adaptation rates, there are mechanisms
that sustain rare strategies in evolutionary dynamics with cyclic
competition. One is spatial structure of contact networks, as
explained in \SEC\ref{sec:introduction}
\cite{Hassell91,Tainaka93,Durrett98TPB,Frean01,Szabo02prl,Szabo04jpa,SzaboVukov04,Wu05pre,Ying07jpa}.
Another important mechanism is mutation.
No strategy dies out 
in the presence of mutation
because mutation decreases the proportions of
major strategies and increases those of rare and absent
strategies. Actually, mutation establishes oscillations
and is a key to altruism in the iterated
Prisoner's Dilemma with cyclically competing three strategies, that
is, ALLC, ALLD, and TFT \cite{Imhof05}.

We have neglected mutation in this work.
Mutation-induced oscillations, which imply
coexistence of multiple strategies,
and oscillations by heterogeneous adaptation rates are consistent.
When mutation creates an oscillation
in otherwise nonoscillatory evolutionary dynamics,
the amplitude of an
oscillation is fairly large because a mutation rate is
typically small.
A large oscillation amplitude implies that, in most of the time, the
proportion of at least one strategy is very small \cite{Imhof05}. With
this case included, heterogeneity lessens the oscillation amplitude so that
the proportion of each strategy does not fluctuate so much in time.

\subsection{Heavily skewed cyclic competition makes coexistence difficult}

To test our theory, we examined two numerical models in
\SEC\ref{sec:numerical}. One is the standard RSP game that directly
corresponds to the analytical model in \SEC\ref{sec:rsp-analytical}.
The other is the public goods game with voluntary participation
\cite{Hauert02sci,Hauert02jtb}. There are other evolutionary games in
which oscillations derived from cyclic competition are observed. Examples
include the Prisoner's Dilemma with a mixture of different memory-one
strategies \cite{Nowak89jamc,Nowak89jtb,Nowak93pnas}, a
seminal model of indirect
reciprocity \cite{Nowak98NAT,Nowak98JTB}, and the iterated Prisoner's
Dilemma with misimplementation of the action \cite{Brandt06jtb}.  In
these models, three or more strategies including altruistic
strategies alternately prosper in infinite well-mixed populations.
By additional numerical simulations, 
these dynamics were examined with $200\le n\le 1000$
and heterogeneous
adaptation rates.  However, it was impossible to stabilize the
oscillatory dynamics with the ranges of parameter values probed
(results not shown). 

\begin{figure}
\begin{center}
\includegraphics[height=2in,width=2in]{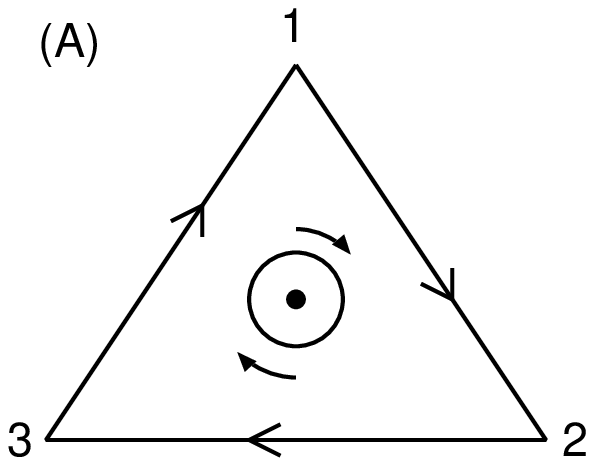}
\includegraphics[height=2in,width=2in]{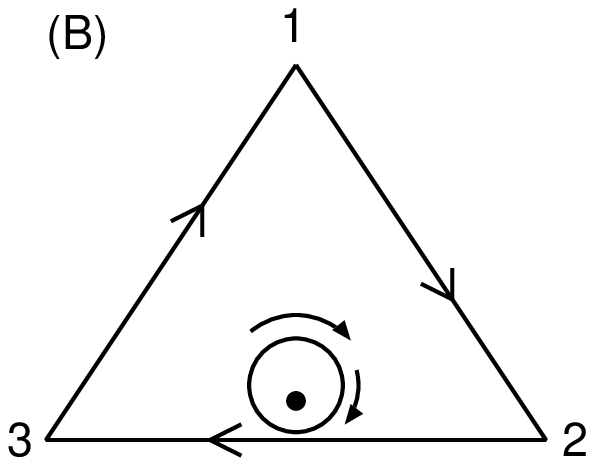}
\caption{Schematic of trajectories of (A) symmetric and (B)
heavily skewed RSP dynamics.}
\label{fig:rsp_triangle}
\end{center}
\end{figure}

We conjecture that this failure is caused by the extreme asymmetry
inherent in these RSP dynamics. To illustrate, suppose a population
with homogeneous adaptation rates. By symmetric RSP dynamics, we mean
those defined by the payoff matrix given in
\EQ(\ref{eq:matrix-nonzerosum3}). The corresponding dynamics of the
population densities are schematically depicted in
\FIG\ref{fig:rsp_triangle}(A) so that the interior equilibrium is
located in the center of the simplex. Each point in the simplex
specifies a density profile of different strategies. A point close to
the corner labeled 1, for example, corresponds to a population that
contains relatively many players with strategy 1.  The symmetric or
quasi symmetric RSP dynamics are often employed in the context of
interacting particle systems as well as evolutionary games
\cite{Tainaka93,Durrett98TPB,Frean01,Szabo04jpa,Reichenbach06pre,Ying07jpa}.
An example of heavily asymmetric, or skewed, RSP dynamics is depicted in
\FIG\ref{fig:rsp_triangle}(B).
To complete a cycle, the trajectory has to
proceed through a narrow canal between the interior equilibrium and the
heteroclinic path on which only strategies 2 and 3 exist (the bottom
line of the triangle). Accordingly, the proportion of strategy 1
is close to zero for some significant time in each cycle.
Such heavily skewed RSP
dynamics are identified in literature \cite{Nowak89jamc,Nowak89jtb}
and may be widely found in evolutionary
games.

Because trajectories for a population of small to intermediate size
accompany considerable fluctuation, it likely hits the heteroclinic
path so that the rare strategy (strategy 1 in
\FIG\ref{fig:rsp_triangle}(B)) perishes. If the location of the
interior equilibrium and trajectories derived from an
infinite-population theory are extremely skewed, oscillatory dynamics
will not be realized in finite populations.  Because the heterogeneous
adaptation rate does not change the position of the equilibrium (see
\SEC\ref{sec:rsp-analytical} and \cite{MasudaRSP}), heterogeneity does
not widen the canal.  Although mutation kicks trajectories that have
fallen onto the heteroclinic path back to the interior of the simplex,
the rare strategy is perpetually subject to extinction.  The
heterogeneity mechanism discovered here will work out for RSP dynamics
that are not extremely skewed.

\section*{Acknowledgments}
We thank Shinsuke Suzuki
for critical reading of the manuscript.

\newpage

\clearpage

\clearpage

\clearpage

\end{document}